\documentclass[twocolumn,showpacs,prl,superscriptaddress,amsmath,amssymb]{revtex4}

\usepackage{graphicx}
\usepackage{bm}

\begin{document}

\title{Dynamics of magnetic domain wall motion after nucleation:\\
Dependence on the wall energy}

\author{K. Fukumoto}
\altaffiliation[Present address: ]{JASRI/SPring-8, 1-1-1, Kouto,
Sayo, Hyogo 679-5198, Japan}

\author{W. Kuch}
\affiliation{
Freie Universit\"at Berlin, Institut f\"ur Experimentalphysik,
Arnimallee 14, D-14195 Berlin, Germany}

\author{J. Vogel}
\affiliation{Laboratoire Louis N\'eel, CNRS, 25 avenue des
Martyrs, B. P. 166, F-38042 Grenoble cedex 9, France}

\author{F. Romanens}
\affiliation{Laboratoire Louis N\'eel, CNRS, 25 avenue des
Martyrs, B. P. 166, F-38042 Grenoble cedex 9, France}

\author{S. Pizzini}
\affiliation{Laboratoire Louis N\'eel, CNRS, 25 avenue des
Martyrs, B. P. 166, F-38042 Grenoble cedex 9, France}

\author{J. Camarero}
\affiliation{Departamento de F\'{\i}sica de la Materia Condensada,
Universidad Aut\'onoma de Madrid, E-28049 Madrid, Spain}

\author{M. Bonfim}
\affiliation{Departamento de Engenharia El\'etrica, Universidade do
Paran\'a, CEP 81531-990, Curitiba, Brazil}

\author{J. Kirschner}
\affiliation{Max-Planck-Institut f\"ur Mikrostrukturphysik,
Weinberg 2, D-06120 Halle, Germany}

\begin{abstract}

The dynamics of magnetic domain wall motion in the FeNi layer of a
FeNi/Al$_2$O$_3$/Co trilayer has been investigated by a
combination of x-ray magnetic circular dichroism, photoelectron
emission microscopy, and a stroboscopic pump--probe technique. The
nucleation of domains and subsequent expansion by domain wall
motion in the FeNi layer during nanosecond-long magnetic field
pulses was observed in the viscous regime up to the Walker limit
field. We attribute an observed delay of domain expansion to the
influence of the domain wall energy that acts against the domain
expansion and that plays an important role when domains are small.

\end{abstract}



\maketitle

Many of the fascinating phenomena in thin film magnetism that
appear promising for new applications in magnetic data storage
involve the use of multilayered magnetic systems, for example
spin-valves \cite{geor} and magnetic-tunnel-junctions
\cite{moodera}. The study of the time dependence of the magnetic
properties in multilayered systems is of high actual interest with
respect to the achievable magnetic response time of such devices,
but also from a fundamental point of view. It has been observed
that the magnetization reversal for hard/soft magnetic trilayers
is dominated by domain wall motion in quasi-static conditions and
by the nucleation of many small domains in fast dynamic conditions
\cite{yan,keiki2}. Such a difference in the reversal mechanism
must obviously be accompanied by dramatic changes in the
energetics, and magnetic domain walls will become much more
influential at faster reversal rates.

The role of domain wall energy on the magnetization reversal is a
long-debated issue.  It has been discussed in the seventies in
connection with bubble domains in perpendicularly magnetized
materials.  Malozemoff and Slonczewski pointed out that the domain
wall energy leads to a wall curvature pressure that tends to
contract an isolated bubble domain, eventually collapsing the
domain when the opposing demagnetizing energy is overcome
\cite{bubble}. Skomski {\it et al.}, in a self-consistent
magnetic-viscosity approach, pointed out a negative velocity for
bubble domain expansion in small fields as consequence of the
domain wall energy, thus confirming the domain collapse
\cite{Skomski}.  More recently, Lyberatos and Ferr\'e studied the
rate of domain growth in an applied magnetic field by Monte Carlo
simulations. The results showed a slowing down of domain wall
motion and the appearance of a fractal shape of the domains due to
pinning effects in the regime of thermally activated domain wall
motion. However, no influence of domain wall energy was found in
the viscous motion regime, in which pinning processes do not
hinder the domain wall propagation \cite{Lyberatos}.
Experimentally, it has been shown that the influence of the domain
wall energy can lead to a straightening of domain walls in the low
velocity creep-like motion regime in the presence of an introduced
defect \cite{1405}, and to a selectively accelerated domain
expansion of such wall motion processes leading to a merging of
domains \cite{wgtr}.

In this Letter we present an experimental investigation of the
influence of the domain wall energy on the magnetization reversal
of the soft magnetic layer in magnetic tunnel junction-like
trilayers. We study the reversal of the soft layer away from the
magnetization direction of the hard layer. In this configuration
the reversal is nucleation-dominated reversal for high fields
\cite{yan,keiki2}, and a strong influence of the domain wall
energy is expected. We use a stroboscopic pump--probe technique
\cite{jan1} to observe in real space the repeated nucleation and
expansion of domains and find that the influence of the domain
wall energy leads to an apparent delay in domain nucleation, which
can amount up to several nanoseconds. This can be well described
by an effective specific domain wall energy exerting a negative
domain wall pressure when domains are small, just after
nucleation, but becoming negligible once domains exceed a certain
size.

In in-plane domains there is no stabilizing demagnetizing energy as
in perpendicular bubble domains, and the domain wall energy $P
\gamma d$ of a domain with perimeter length $P$ in a film of
thickness $d$ \cite{Skomski} can be
described simply by an effective field $H_{wall} = \gamma
\lambda /(4 \mu_0 M_S P)$ acting against the external field
$H_{ext}$ \cite{bubble}.  Here $\gamma$ is an effective domain
wall energy averaged over the perimeter of the domain, and
$\lambda$ is a geometric factor
describing the ratio between area and squared perimeter of the domain
($\lambda = 4 \pi$ for a circular domain).
Only if $H_{wall}$ is small enough so that the
corresponding wall velocity can be neglected, domains are stable in
static conditions.  In our case both, the domain wall
field and the effective field from the coupling between the two
ferromagnetic layers in a trilayered structure, $H_{copl}$,
oppose the externally applied field, so that the total effective field
is
\begin{equation}
    H_{tot} = H_{ext} - H_{copl} - \frac{\gamma \lambda}
    {4 \mu_0 M_S P},
    \label{neweq}
\end{equation}

and thus depends on $P$.  We consider as measurable quantity the
speed of perimeter expansion, $dP/dt$, which in turn will be a
function of $H_{tot}$.  We show that the resulting inhomogeneous
non-linear differential equation for $P$ can well describe the
observed apparent delay in reversible domain nucleation using a
fixed value for $\gamma \lambda$, thus revealing the manifestation
of the domain wall energy in the reversal dynamics.

The sample used in this study was a 4 nm Fe$_{20}$Ni$_{80}$/2.6 nm
Al$_{2}$O$_{3}$/7 nm Co trilayer grown by RF sputtering on the
step-bunched surface of a 6$^{\circ}$ miscut Si(111) crystal.
Underneath the Co layer, 3 nm of CoO has been deposited to
increase the coercivity of the Co layer. The FeNi layer has been
capped by 3 nm of Al to prevent oxidation. Details of the thermal
treatment leading to the creation of step-bunches are given in
Ref.\ \cite{encinas}. According to quasi-static longitudinal Kerr
effect experiments, the sample features uniaxial anisotropy along
the step-bunches. The coercivities are 1 mT for the FeNi and 4.3
mT for the Co layers, and a N\'eel type coupling field $\mu_0
H_{copl}$ of 1.2 mT favors parallel alignment. Details of similar
types of samples are shown in Refs.
\cite{yan,jan1,janani,wgtr,keiki2}.

The experiment was performed using a combination of x-ray magnetic
circular dichroism (XMCD), photoelectron emission microscopy
(PEEM), and a stroboscopic pump--probe technique \cite{jan1}.
Magnetic field pulses are periodically applied at 312.5 kHz by a
coil made of Cu foil wrapped around the sample.  The magnetic
response is stroboscopically probed by XMCD-PEEM with the same
frequency at a fixed phase delay with respect to the magnetic
field pulses.  The time resolution is given by the x-ray pulse
width and the jitter of the electronics, and is estimated as 100
ps. The domain wall motion in the two ferromagnetic (FM) layers
was observed separately by tuning the photon energy of circularly
polarized x-rays to the Fe and Co $L_3$ absorption edges for
addressing the FeNi and Co layer, respectively. Measurements were
performed at the UE56/2-PGM2 beamline at BESSY in Berlin.

\begin{figure}[tbp]
\includegraphics*[bb= 175 378 395 521]{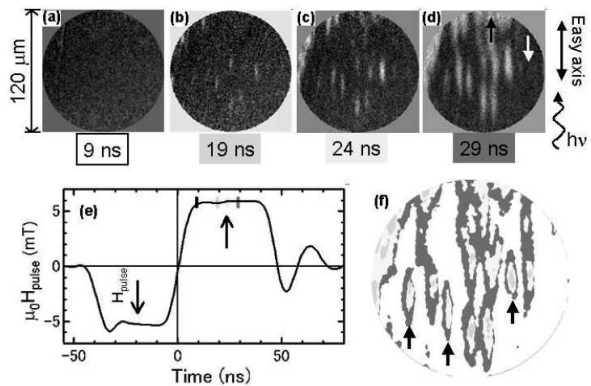}
\caption{Bipolar magnetic field pulses (e) were used to reverse the
magnetization in part of the FeNi layer. The amplitude of the field
pulses was 5.80 mT on the plateaus of both the positive and negative
part of the pulses. The magnetic domain structure in the FeNi layer
at different delay times is shown in (a--d), where the photon energy
of circularly polarized x-rays was tuned to the Fe $L_3$ absorption
edge. White domains in (b) to (d) are discretized and superimposed
to (f). Different shades of grey in (f) indicate different delay
times. Images taken at the Co-L$_{3}$ edge confirmed that no
magnetization changes took place in the Co layer. } \label{f1_elli}
\end{figure}

In Fig.\ \ref{f1_elli} we show images of the nucleation of domains
and their subsequent expansion in the FeNi layer, observed at an
amplitude of the bipolar field pulses of 5.8 mT. The pulse shape
is shown in Fig.\ \ref{f1_elli} (e). The positive and negative
parts have the same amplitude (5.8 mT) and length (47 ns). Figs.\
\ref{f1_elli} (a), (b), (c), and (d) show the magnetic domain
structures in the FeNi layer at different pump--probe delay times,
9, 19, 24 and 29 ns, respectively.  The easy axis is parallel to
the vertical direction in the images, and the black and white
contrasts indicate the magnetization pointing down and up,
respectively. The direction of the incoming x-rays with
30$^{\circ}$ grazing angle to the surface is also indicated in
(d). The field of view is 120 $\mu$m.

Before applying the pulses, the two FM layers were magnetically
saturated in the same direction by a 15 mT field pulse of 3 ms
duration. The direction is indicated by the white arrow in Fig.\
\ref{f1_elli} (d), and corresponds to black contrast. During the
pump--probe experiments, the first, negative part of the pulse is
in the same direction as the saturation, and was used to obtain
the same starting condition for each pulse. The second, positive
part is in the opposite direction (see (e)). The magnetization in
the FeNi layer starts to reverse into the white direction during
this second part of the pulse.  The start of the positive part of
the pulse is set to zero time. The plateau of the pulse starts at
9 ns, at which time one can see only some small fuzzy white
splotches, probably due to magnetic domains that are blocked on
surface defects (a). However, they do not expand with time, and
are thus not regarded in the analysis. Some white domains become
visible in (b), and, as time goes by, they get larger (c and d).
To get sufficient statistics, millions of pulses were applied to
acquire each image.

The areas of white domains in Fig.\ \ref{f1_elli} (b), (c), and (d) are
superimposed into (f) with different shades of grey, middle grey,
light grey, and dark grey, respectively. The expansion of the
domains by propagation of the domain walls is clearly seen. In
this experiment, the magnetization of the Co layer was not
affected by the field pulses, i.e., it always showed the black
contrast.

\begin{figure}[tbp]
\includegraphics*[bb= 185 320 406 480]{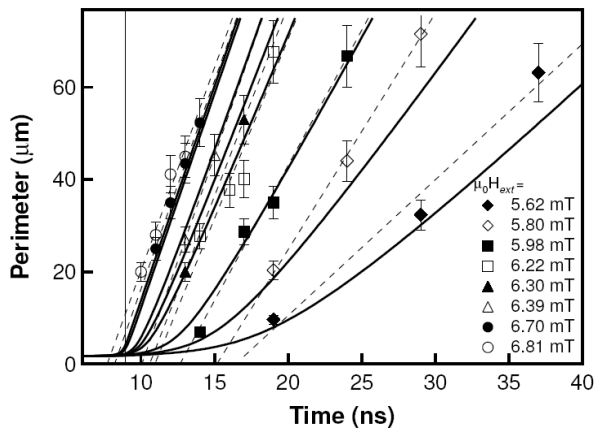}
\caption{Increase of the domain perimeter vs.\ time during the
plateau of the positive part of the field pulses for several field
amplitudes.  Broken lines are linear fits to the data.  A simulation
of the perimeter extension with time is shown by solid curves. }
\label{f2_perimeter}
\end{figure}

In Fig.\ \ref{f2_perimeter}, the average perimeter of the three
domains indicated by arrows in Fig.\ \ref{f1_elli} (f) is plotted as
a function of time for each amplitudes $H_{pulse}$ of the field.
$\mu_0 H_{ext}$ is ranging from 5.62 to 6.81 mT. The perimeter $P$
is measured by an ellipse-fit to the domains with an error of about
10\%.  The three domains were nucleated for all amplitudes of the
field pulses, and it was possible to measure $P$ before they merged
with other surrounding domains. The start of the plateau is
indicated by a vertical line in Fig.\ \ref{f2_perimeter}.  Linear
fits to the data (broken lines) indicate that the measured values
for $P$ extended linearly with time for each field amplitude.
However, the intersections to the time axis of these lines do not
come to the same point, but show an apparent delay in the appearance
of the domains, which amounts up to 17 ns for $H_{ext} = 5.62$ mT.

\begin{figure}[tbp]
\includegraphics*[bb= 179 339 399 481]{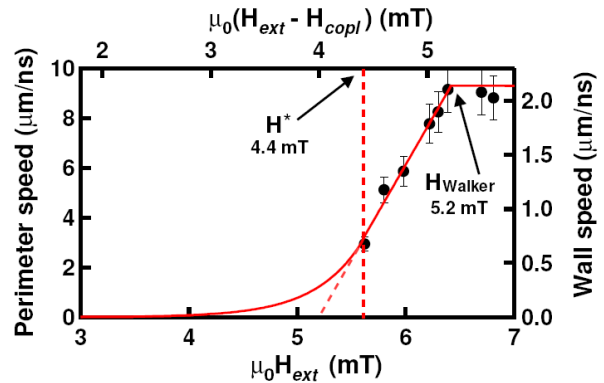}
\caption{Speed of perimeter extension vs.\ $\mu_0 H_{ext}$ (bottom
axis) and $\mu_0 (H_{ext}-H_{copl})$ (top axis). The linear fit to
six of the data points ranging from 4.4 mT to 5.2 mT of $\mu_0
(H_{ext} - H_{copl})$ indicates that the extension of the perimeter
(i.e., the wall motion) is in the viscous regime.  The exponential
part at lower fields ($\mu_0 H_{tot} < 4.4$ mT) is drawn using Eq.\
\ref{eq3}. In the higher field region, above $H_{Walker}$, the
perimeter speed is assumed to be constant (9.3 $\mu$m/ns). }
\label{f3_speed}
\end{figure}

In Fig. \ref{f3_speed}, the slope of linear fits to the
experimental data of Fig. \ref{f2_perimeter}, i.e., the speed of
perimeter expansion for large domains, is plotted vs.\ $\mu_0
H_{ext}$ (bottom axis) and vs.\ $\mu_0 (H_{ext} - H_{copl})$ (top
axis). In the range of $\mu_0 H_{ext}$ between 5.6 and 6.4 mT,
$dP/dt$ increases linearly with field, meaning that the wall
motion is in the viscous regime in this field range.  A mobility
of {\it perimeter extension}, $\mu$, of 7\,800 m/(s mT) $\pm$10\%
was obtained from the slope of a linear fit. The two data points
with the highest amplitudes of the field pulses (6.70 mT and 6.81
mT) are not included in the linear fit, since the velocity seems
to reach saturation at around 5.2 mT of $\mu_0
(H_{ext}-H_{copl})$. This saturation of velocity is due to the
modification of the spin structure inside the domain wall above
the so-called Walker limit field \cite{walker,yuan}.

For the analysis according to Eq.\ \ref{neweq} we use the data of
Fig.\ \ref{f3_speed} to extract the dependence of $dP/dt$ on
$H_{tot}$, where we assume $H_{tot} \approx H_{ext}-H_{copl}$ for
large domains. The negative contribution of $H_{wall}$ to $H_{tot}$
for small domains makes it necessary to extend the velocity data to
the lower field range for the numerical evaluation of $P(t)$.  We
assumed thermally activated wall motion for lower $H_{tot}$ with
perimeter extension velocity
\begin{eqnarray}
v_P = v_{P,0} ~ exp(\frac{\mu_0 M_S H_{tot} V_B -
E_p}{k_B T}), \label{eq3}
\end{eqnarray}
where $V_B$ and $E_p$ are the Barkhausen volume and the barrier
energy, respectively \cite{ferre,kiril}, and $\mu_0 M_S$ = 1 T for
FeNi. To provide a smooth transition to the experimentally
determined regime of viscous motion at $H^*$ = 4.4 mT, values of
$V_B$ = ($1.2 \pm 0.1) \times 10^{-23}$ m$^3$ and $E_p$ = (2.6
$\pm$ 0.4) $\times$ 10$^{-20}$ J  \cite{yan} have to be used
(solid line in Fig. 3), while $v_{P,0}$ of 55 nm/ns
\cite{remark3}. At $H_{tot} > 5.2$ mT, a constant perimeter
expansion speed of 9.3 $\mu$m/ns was assumed in order to match the
two data points at 6.70 and 6.81 mT.

The perimeter extension was numerically evaluated according to Eq.
\ref{neweq} with 1 ps time increments \cite{remark} and is plotted
in Fig. \ref{f2_perimeter} by solid curves. The best agreement was
obtained for $P(t=0)$ = 1.9 $\mu$m and $\gamma\lambda$ = 9
mJ/m$^2$.  The simulation fits quite well to the experimental
data. $P(t=0)$ and $\gamma \lambda$ are found to be correlated by
$\gamma \lambda = 3 \times 10^4~{\rm{(mJ/m^3)}}~P(t=0) +
3.2~{\rm{(mJ/m^2)}}$, and no significant worsening of the fit is
observed for values of $\gamma \lambda$ between 7 and 11 mJ/m$^2$,
if $P(t=0)$ is changed accordingly and kept within 3\% of the
respective value.  It is seen that a deviation from the linear
behavior occurs at the initial stages of domain expansion, thus
leading to the observed apparent delay. For higher $\mu_0 H_{ext}$
(6.70 mT and 6.81 mT), $P$ starts to expand already on the rising
slope of the pulse before 9 ns, as was also observed in the
experiment.

The above simulation assumed that there is no real delay in domain
nucleation. Since the grey scale contrasts obtained in a static
experiment were identical to the ones of the images obtained in
the above experiment, the reproducibility of domain nucleation is
close to 100\%. Once the domains were visible, we did not observe
an increase of the number of domains during the field pulses. This
indicates that the domains were always nucleated as soon as the
pulse field reached a certain threshold.  The observation of a
reproducible nucleation of domains at certain locations of the
sample means that topographic features act as nucleation centers,
and that the nucleation process is not thermally activated.
Reproducible nucleation takes place at those positions where the
local nucleation barrier is lower than the effective field. The
increase of the number of nucleated domains with increasing pulse
height indicates therefore a distribution of barrier heights.

We have assumed a constant elliptical shape for the domains with
aspect ratio 1:4, thus $\lambda \approx 24$.  The perimeter speed
can be converted into a domain wall speed along the easy axis of
magnetization.  This is plotted on the right axis of Fig.\
\ref{f2_perimeter}. The mobility of viscous domain wall motion
along the easy axis is then 1\,600 m/(s mT) $\pm10\%$. The initial
$P$ of 1.9 $\mu$m means that the width of the nucleation centers
is around 0.19 $\mu$m, which is too small to be seen in the
images, the lateral resolution of which was set to $\approx 1$
$\mu$m for these measurements. The maximum speed of 2000 m/s along
the easy axis is an order of magnitude larger that the value found
for narrow permalloy lines using similar field values, indicating
that it might be possible to increase this speed by an appropriate
shaping of the wires \cite{beach}.

The Walker limit field is given by $\mu_0 H_{Walker}$ = 1/2$\alpha
\mu_0M_S$ \cite{walker}. For Fe$_{20}$Ni$_{80}$, a value $\alpha =
0.01$ is often used in the literature for the damping constant
\cite{niba,ingv,keiki2}. With this value, $\mu_0 H_{Walker}$ would
be 5 mT.  This is in good agreement with our experimentally
obtained value of 5.2 mT.

The constant effective domain wall energy $\gamma = 0.38$ mJ/m$^2$
that was used for the simulation includes exchange and anisotropy
energy inside the wall, magnetostatic contributions from the sharp
ends of the elliptic domains, as well as repulsion of the long
sections of the magnetic domain walls \cite{mag}.  A refined model
should take into account the dependence of $\gamma$ on $P$. Since
in the simulation $\gamma$ contributes mainly when domains are
small, deviations in the shape of the domains may also play a role
when comparing the value for $\gamma$ to an estimate for the wall
energy of an isolated N\'eel type wall, which leads to a much
higher value of 2.3 mJ/m$^2$ \cite{remark2}.

In conclusion, the reproducible nucleation of domains in the soft
layer of a tunnel junction-like trilayer and the subsequent domain
growth by domain wall propagation in the viscous regime and up to
the Walker limit field is significantly influenced by the energy
of magnetic domain walls.  This leads to an apparent delay of
domain expansion.  A faster magnetization reversal, which is
required for the coming magnetic recording devices, is drastically
hindered by this magnetic domain wall energy when magnetic domains
are small. By analogy, the techniques and results shown in this
work are equally applicable to other nucleation and growth
phenomena like crystallographic phase changes, superconducting
transitions, and solidification theory.

We thank F.\ Helbig, B.\ Zada, W.\ Mahler, and G.\ Meyer for
technical support, F.\ Petroff and A.\ Vaur$\rm{\grave{e}}$s for
sample preparation.  Financial support by BMBF (no.\ 05KS1EFA6),
EU (BESSY-EC-HPRI Contract No.\ HPRI-1999-CT-00028) and
Laboratoire Europ\'een Associ\'e (LEA) "Mesomag" is gratefully
acknowledged.


\begin{thebibliography}{99}
\bibitem{remark}
    For the simulation, the pulse shape was simplified
such that the field increases linearly with time during the
rise-time, the slope thus being $\mu_0 H_{ext}/9$ ns. After 9 ns,
the field is assumed constant up to the end of the plateau.
\end{thebibliography}


\end{document}